\documentclass[12pt,a4paper]{article}

\usepackage{latexsym} 
\title{Time dependence of $c$ and its concomitants}

\author{Ll.\ Bel\\
\emph{Fisika Teorikoa, Euskal Herriko Unibertsitatea}, \\
\emph{P.K. 644, 48080 Bilbo, Spain} \\
\emph{e-mail: wtpbedil@lg.ehu.es}
}

\begin{document}
\maketitle

\begin{abstract} 

As we showed in a preceding arXiv:gr-qc Einstein equations, conveniently
written, provide the more orthodox and simple description of
cosmological models with a time dependent speed of light $c$. We derive
here the concomitant dependence of the electric permittivity $\epsilon$,
the magnetic permeability $\mu$, the unit of charge $e$, Plank's
constant $h$, under the assumption of the constancy of the fine
structure constant $\alpha$, and the masses of elementary particles $m$.
As a consequence of these concomitant dependences on time they remain
constant their ratios $e/m$ as well as their Compton wave length
$\lambda_c$ and their classical radius $r_0$. 

\end{abstract}

\section{Introduction}

To say that light rays are null geodesics of some space-time metric is
an intrinsic statement, but any statement about the speed of light is
necesserally relative to a frame of reference, as soon as we escape from
the semantic tautology that consists in saying that the speed of light
is $c$ because the universal constant $c$ is called the speed of
light... To say that the speed of propagation of light from a point A to
a point B and back is some given value $W=D/\triangle T$ requires that
we define the time interval $\triangle T$ as proper time along the
trajectory of A, and this does not raise any problem, but it requires
also that we define the space distance between the world-lines of A and
B and this is subject to debate. More generally this requires to define
the time-like congruence of world lines which will be the first
ingredient of the frame of reference where the measures of distance are made. 
To say
that the speed of propagation of light from a point A to a point B is
some given value $V=D/(T_B-T_A)$ further requires to specify a
synchronization of time, i.e. the second ingredient of any frame of
reference, between the world-line of A and that of B to make sense of
the denominator of this formula.

There is a point of view that allows to speak about the speed of
light without falling into any fundamental metrology problem. Let us
assume that we know of a theory which we know how to test or that has
been useful to describe scenarios we are interested in. Let us assume
that the formalism of this theory includes a constant, say $c_0$, with
the dimensions of a velocity and somehow we have some justification to
suspect that this constant could in fact be a function, possibly a new
unknown of a new theory grafted to the main one. Then it might be
legitimate, even before the operational meaning of this function has
been clarified, to make the effort to discuss the new theory if we are able
to make sense of its new predictions or the new scenarios that follow
from it. At least temporarily, because sooner or later we shall need to
clarify what we mean by frame of reference, space, time and
velocity\,\footnote{Our point of view has been described in 
\cite{Frames} and references therein} 

The idea that the speed of light could depend on gravity was one of
those that Einstein had in mind since his 1907 paper when he started
thinking about a generalization of Special relativity, and later on in
his 1911 and 1912 papers. Einstein changed his mind in several ocasions
and he did it loudly, as when he decided that the field equations had to
be covariant after having argued earnestly that covariance was a
physical nonsense\,\footnote{In our opinion it is still a mistake to
confuse covariance with invariance under a dynamical group}, or when he
regreted to have used a cosmological constant to propose a static model
of the Universe. But, to our knowledge, he never made public that he has
changed his mind about the speed of light: he just ceased to write about
it. To most people, this idea is now anathema and has been abandoned in
favor of the hypothesis that the speed of light is the same constant
whatever the location and time and whatever the frame of reference, a
concept which is often used without care and identified with any system
of coordinates. 

For us the time and location dependence of the speed of light is an
inescapable consequence of a theory of frames of reference that
implements the axiom of free mobility \cite{Frames}. This paper is
nevertheless independent from these general considerations and is
self-contained.\,\footnote{For a recent review on different Variable
Speed of Light theories in cosmology read \cite{Magueijo} and references
therein. These theories compete with inflationary ones to cure problems
of standard cosmological models}

In Sect.\ 2 we review the main idea about the time dependence $c(T)$ of
the speed light for Robertson-Walker cosmological models. In the following
sections, with an increasing level of speculation, we derive models for
the concomitant variation of $e(T)$, the electronic charge, of $\epsilon(T)$,
the electric permittivity of the cosmological medium, of $\mu(T)$, its
magnetic permeabiliy, of $h(T)$, Plank's constant and of $m(T)$, the masses of
elementary particles. We assume in the process the constancy of
$\alpha$, the fine structure constant, and $G$, Newton's constant\,\footnote{
Observational constraints on the variation of $\alpha$ and $G$ are discussed 
in \cite{Uzan}}. 

\section{$c$ is a function of time}

In a recent preprint, \cite{Locexpan}, we considered the general Robertson-Walker line-element  
written as follows:

\begin{equation}
\label {1.1}
dS^2=-dT^2+\frac{1}{c^2(T)}\left(\frac{dR^2}{1-
kR^2}+R^2d\Omega^2\right), \quad
d\Omega^2=d\theta^2+\sin^2\theta
d\varphi^2
\end{equation}

with $c(T)$ having the dimensions of velocity and $k$, the space
constant curvature, those of inverse square length. The idea that this
rewriting makes explicit is that $c(T)$ is a varying speed of light and
that, what is usually called the scale factor is:

\begin{equation}
\label {1.2}
F(T)=c_0/c(T)
\end{equation}
where $c_0$ is the measured speed of light at any reference time $T_0$,
and thus $F(T)$ can be interpreted as the corresponding varying
refractive index of the cosmological medium with $F(T_0)=1$ at the time
when $c(T_0)=c_0$. 

Notice that from this point of view $c_0$ ceases to be a universal
cosmological constant to become a local constant to be measured at
any time that one wishes to describe local physics.

The field equations are:

\begin{equation}
\label{1.3}
S_{\alpha\beta}+\Lambda g_{\alpha\beta}=8\pi GT_{\alpha\beta}
\end{equation}
where $S_{\alpha\beta}$ is the Einstein tensor of the line-element
(\ref{1.1}), $\Lambda$ is the proper global cosmological constant of
the model with dimensions ${\hbox{T}}^{-2}$ and $T_{\alpha\beta}$ is:

\begin{equation}
\label{1.4}
T_{\alpha\beta}=\rho(T)u_\alpha u_\beta 
+\frac{P(T)}{c(T)^2}(g_{\alpha\beta}+u_\alpha u_\beta)
\end{equation}
with $u_0=-1$ and $u_i=0$. $G$ is Newton's constant and it is assumed
to be indeed constant. $\rho(T)$ is mass density and $P(T)$ is pressure.
Eqs. (\ref{1.3}) are then the usual equations with a small but important
difference: the function $c(T)$ in the r-h-s of (\ref{1.4}) is a
replacement for a constant $c_0$, considered to be a universal constant.
This is 
the single presence of the speed of light in the
r-h-s. In other words, there is no need for Einstein's constant.

The explicit Einstein's equations (\ref{1.3}) reduce to the following
two:

\begin{equation}
\label{1.5}
3\frac{kc(T)^4+\dot c(T)^2}{c(T)^2}=8\pi G\rho(T)+\Lambda
\end{equation}   
 
\begin{equation}
\label{1.6}
\frac{kc(T)^4+5\dot c(T)^2-2c(T)\ddot c(T)}{c(T)^2}=-8\pi
G\frac{P(T)}{c(T)^2}+\Lambda
\end{equation} 
where a dot means a derivative of a function with respect to $T$.

Eqs. (\ref{1.3}) remain covariant keeping in mind that the metric
coefficients $g_{\alpha\beta}$ in both sides transform as a covariant
tensor, that $u_\alpha$ transform as a covariant vector, and that
$\rho(T)$, $p(T)$ and $c(T)$ in the r-h-s must be transformed as
scalars. From the usual point of view, i.e. writing in the r-h-s $c_0$
instead of $c(T)$ the Eqs. (\ref{1.3}) are also obviously manifestly
covariant but the physics that they describe is local in time and
different from the physics that we consider to describe the cosmological
model as a whole.

\section{$\epsilon$, $\mu$ and the unit of charge $e$ are functions of time}

Assuming, as we do, that the cosmological medium behaves as a linear dielectric, 
and that the speed of light is a function $c(T)$, means equivalently to assume that
either the electric permittivity
$\epsilon(T)$ or the magnetic permeability $\mu(T)$, or both are also functions of time, 
the relationship between these three quantities being: 

\begin{equation}
\label {1.7}
c(T)=1/\sqrt{\epsilon(T)\mu(T)} \quad \hbox{or} 
\quad F(T)=c_0\sqrt{\epsilon(T)\mu(T)}
\end{equation}   

We know, (\cite{Pham} and \cite{Synge}), that under these circumstances
the space-time trajectories of light rays are not the null geodesics of
(ref(1.1)) with metric $g_{\alpha\beta}$ but those of the metric:

\begin{equation}
\label {1.9}
\bar g_{\alpha\beta}=g_{\alpha\beta}+(1-F(T)^{-2})u_\alpha u_\beta
\end{equation} 

We know also that under the same circumstances the Maxwell
equations are, if no charges nor currents are present:

\begin{equation}
\label {1.10}
\partial_\alpha F_{\beta\gamma}+\partial_\beta F_{\gamma\alpha}+
\partial_\gamma F_{\alpha\beta}=0
\end{equation} 

\begin{equation}
\label {1.11}
\partial_\alpha(\sqrt{-g}G^{\alpha\beta})=0 , 
\quad \sqrt{-g}=\frac{F(T)^3R^2\sin\theta}{c(T)^3\sqrt{1-kR^2}}
\end{equation}
where:

\begin{equation}
\label {1.12}
G^{\alpha\beta}=\frac{1}{\mu}\bar g^{\alpha\rho}\bar g^{\beta\sigma}F_{\rho\sigma}
\end{equation}
with:

\begin{equation}
\label {1.13}
\bar g^{\alpha\rho}\bar g_{\rho\beta}=\delta^\alpha_\beta
\end{equation}

If the line-element were Minkowski's metric, i.e. if $c(T)=c_0$ were
constant and $k$ were zero,  the electromgnetic field $F_{\alpha\beta}$ of a 
point monopole charge, say $e_0$, would be:

\begin{equation}
\label {1.13.0}
F_{10}(T,R)=\frac{e_0}{4\pi\epsilon_0 R^2}, \quad F_{20}=F_{30}=0, \quad F_{ij}=0
\end{equation} 
where ($i,j,..=1,2,3$).

We want to find the corresponding solution of Maxwell equations under
the assumption that $\epsilon(T)$, $\mu(T)$ as well as $e(T)$ may be
functions of T, and that $k\not=0$ may change the $R$ dependence. More
precisely we lok for a solution of the following form:

\begin{equation}
\label {1.14}
F_{10}(T,R)=\frac{e(T)}{4\pi\epsilon(T)D(R)^2}, \quad F_{20}=F_{30}=0, \quad F_{ij}=0
\end{equation}
This field already satisfies Eqs. (\ref{1.10}).
From (\ref{1.9}) it follows that the non zero components of
$\bar g_{\alpha\beta}$ are:

\begin{equation}
\label {1.15}
\bar g_{00}=-F(T)^{-2}, \quad \bar g_{11}=c(T)^{-2}(1-kR^2)^{-1},
\quad \bar g_{22}=\bar g_{33}=\sin^{-2}\theta=c(T)^{-2}R^2
\end{equation} 
and from (\ref{1.7}), (\ref{1.12}) and (\ref{1.13}) it follows that the non zero
component of $G^{\alpha\beta}$, which reduces to $-e_0/(4\pi\epsilon_0 R^2)$ in 
Minkowski's space-time, is:

\begin{equation}
\label {1.16}
G^{10}=-(1-kR^2)\frac{c_0^4e(T)}{4\pi F(T)^2D(R)^2}
\end{equation}
Giving to $\beta$ the value $0$, Eq. (\ref{1.11}) gives:

\begin{equation}
\label {1.17}
D(R)^2= R^2\sqrt{1-kR^2}
\end{equation} 
and giving to $\beta$ the value $1$, Eq. (\ref{1.11}) gives:

\begin{equation}
\label {1.18}
e(T)=e_0F(T)^{-1}
\end{equation}
$e_0$ being the value of e(T) at the time of reference when
$c(T)=c_0$.

From (\ref{1.15}) it follows that:

\begin{equation}
\label {1.19}
\sqrt{-\bar g}=F^{-1}\sqrt{-g}, \quad  
\bar g=\det(\bar g_{\alpha\beta}),
\end{equation}
therefore, using (\ref{1.12}), Eq. (\ref{1.11}) can be written:

\begin{equation}
\label {1.20}
\partial_\alpha(\sqrt{-\bar g}(F/\mu)\bar F^{\alpha\beta})=0, \quad
\bar F^{\alpha\beta}=g^{\alpha\rho}g^{\beta\sigma}F_{\rho\sigma}
\end{equation}
Although not compelling it is attractive to make the further
assumption:

\begin{equation}
\label {1.21}
\mu(T)=\mu_0 F(T)
\end{equation} 
$\mu_0$ being the value of $\mu(T)$ when $c(T)=c_0$.
This amounts to make an assumption on the cosmological medium, namely
that $(F_{\rho\sigma}, \bar F^{\alpha\beta})$ will behave as it does the
electromagnetic field in a perfect vacuum in a space-time
with metric (\ref{1.9}). This allows, {\it mutatis mutandis} to refer,
if necessary, to electromagnetic physics as usual.

From (\ref{1.7}) and our new assumption (\ref{1.21}) we have then:

\begin{equation}
\label {1.22}
\epsilon(T)=\epsilon_0F(T)
\end{equation}
$\epsilon_0$ being the value of $\epsilon(T)$ when $c(T)=c_0$. This
completes the determination of the electromagnetic concomitants to a
varying speed of light in cosmology\,\footnote{An {\it Ad hoc} field
theoretical theory describing a space-time dependence $e(x^\alpha)$ was
proposed in \cite{Bekenstein}}. 

\section{$h$ is a function of time and $\alpha$ it is not}

Although both the initial interpretation of Sect.~2 which led to
(\ref{1.7}) and the assumption in Sect.~3 that led to (\ref{1.22}) are
speculative we feel that they are sufficiently justified to deserve
a calm evaluation. The considerations that follow on the contrary
are at this time very loosely motivated and are given as possible
canditates to a time variation of some other fundamental quantities, 
according to our personal ingenuity.   	 

Let us consider the fine structure constant:

\begin{equation}
\label {2.1}
\alpha(T)=\frac{e(T)^2}{2\epsilon(T)h(T)c(T)^2}
\end{equation}
where we have assumed that Plank's constant $h$ can be a function of
time. From the preceding 
section we get:

\begin{equation}
\label {1.23}
\frac{\dot\alpha(T)}{\alpha(T)}=
-2H(T)-\frac{\dot h(T)}{h(T)}
\end{equation}
where $H(T)=\dot F(T)/F(T)$ is the Hubble's constant. If we assume that $h$ does not depend
on $T$ and we consider that $H_0$, the Hubble's constant now is of
the order of $70$ Km/s/Mpc we obtain:

\begin{equation}
\label {1.24}
\frac{\dot\alpha_0}{\alpha_0}=7.1\, 10^{-11} \hbox{\ yr}^{-1}
\end{equation} 
which is a ridiculous large figure compared with present estimates (\cite{Webb}):

\begin{equation}
\label {1.25}
\frac{\dot\alpha_0}{\alpha_0}\approx 2.2\, 10^{-16} \hbox{\ yr}^{-1}
\end{equation}
Therefore we can accept either that the Plank's constant is indeed
constant and then our model of varying speed of light is grossly inconsistent
with observations or that (\ref{1.25}) supports, for the time being, a
variation of the form\,\footnote{In \cite{Magueijo} 
a functional dependence $h(c)$ was discussed} :

\begin{equation}
\label {1.26}
h(T)=h_0F(T)^{-2}
\end{equation}  
i.e.: that the fine structure constant is indeed constant, 
with a precision of $\approx$ 1 part in 10${}^5$.

\section{The masses of elementary particles}

Our next step of speculation concerns the time dependence of the masses
of elementary particles. in (\cite{Locexpan}) we proposed a model to
describe the time-dependent local dynamics of an otherwise isolated 
gravitational system influenced by the global or local cosmology. This, 
in the particular case of a spherically symmetric compact mass $m_0$, led 
to the consideration, at the lowest non trivial approximation, 
to describe this dynamics by the Lagrangian:  

\begin{equation}
\label {1.27.0}
L=\frac12 F(T)^2 
({\dot R}^2+R^2({\dot\theta}^2+\sin^2\theta{\dot\varphi}^2))
-\frac{Gm_0}{R}F(T)^{-1}
\end{equation} 
In (\cite{Locexpan}) we mentioned that the last term in this formula 
suggested that $G$ was a time dependent constant. This is actually in 
contradiction with the fact that the constancy of $G$ is essential 
to give an unambiguous meaning to (\ref{1.3}). Therefore we claim now 
that a better interpretation is to propose the following time-variation 
of the mass $m(T)$ of elementary particles:

\begin{equation}
\label {1.27}
m(T)=m_0F(T)^{-1} 
\end{equation}    
It is rather reassuring the fact that this implies that $e(T)/m(T)$ remains a 
true constant as well as the Compton wave-length and the classical radius of 
elementary particles:

\begin{equation}
\label {1.27.1}
\lambda_c=\frac{h(T)}{m(T)c(T)}=\lambda_c(T_0), \quad 
r_0=\frac{e(T)^2}{4\pi\epsilon(T)m(T)c(T)^2}=r_0(T_0) 
\end{equation} 

We point out below another nice consequence of this choice (\ref{1.27}). 
Let us consider the Klein-Gordon equation for a massive scalar field:

\begin{equation}
\label {1.28}
\left(\Box-\frac{m(T)^2c(T)^4}{\hbar(T)^2}\right)\Phi=0
\end{equation}  

Using the line-element (\ref{1.1}) this equation becomes:

\begin{equation}
\label {1.29}
\left(-\partial_T^2-3H(T)\partial_T+c(T)^2\bar\triangle
-\frac{m(T)^2c(T)^4}{\hbar(T)^2}\right)\Phi=0
\end{equation}  
$\bar\triangle$ being the Laplacian of the space metric:

\begin{equation}
\label {1.33}
d\bar S^2=\frac{dR^2}{1-kR^2}+R^2d\Omega^2
\end{equation}
Using (\ref{1.2}), (\ref{1.26}) and (\ref{1.27}), (\ref{1.29}) becomes:

\begin{equation}
\label {1.30}
\left(-\partial_T^2-3H(T)\partial_T+c(T)^2(\bar\triangle
-\frac{m_0^2c_0^2}{\hbar_0^2})\right)\Phi=0.
\end{equation}
$n_I$ being the collection of eigen-values of the Laplacian
operator $\bar\triangle$ and $J_I(X^i)$ being the corresponding 
eigen-functions\,\footnote{for details see \cite{BD}, \cite{Fulling}
and references therein}:

\begin{equation}
\label {1.30.1}
\triangle J_I(x^i)=n_IJ_I(x^i)  
\end{equation} 
the modes of the scalar field $\Phi(x^\alpha)$ that define the quantum
vacuum are a complete set of solutions of this equation 
of the following form:

\begin{equation}
\label {1.32}
\varphi_I(T,x^i)=u^\pm_I(T) (T)J_I(X^i),  
\end{equation}
which is the union of two sets of 
positive and negative energy\,\footnote{See also \cite{Modes}}. 

Eq. (\ref{1.30}) has to be compared with the most commonly used:

\begin{equation}
\label {1.31}
\left(-\partial_T^2-3H(T)\partial_T+F(T)^{-2}c_0^2\bar\triangle
-\frac{m_0^2c_0^2}{\hbar_0^2}\right)\Phi=0
\end{equation}
where one assumes that $c_0,\ h_0$ and $m_0$ are universal constants.
The comparison shows that using Eq. (\ref{1.30}) a non zero mass $m_0$
just shifts by a constant the eigen-values $n_I$ but keeps invariant the
functional dependence of the modes with respect to the space-time
variables. In other words once the modes for $m_0=0$ are known the modes
with $m_0 \not= 0$ are trivially deduced from them. 

\section{Conclusion}

Assuming that a fundamental constant, say $c$, is in fact a function of
time requires a close examination of all the other constants, here
called concomitants, that with $c$ enter in expressions that have a
physical meaning. This paper has been an attempt to present a
coherent scheme to fulfil such requirement. The points 
that we want to remind or emphasize are the following: 
 
1.- General relativity and Robertson-Walker models, conveniently
interpreted, describe a time-dependence of the speed of light. There is
no need to graft any new field theoretical theory to have a varying
speed of light.

Our approach is a simple example of a more general one, based on a full
fledged theory of frames of reference, that considers the speed of light
in a round trip to be in general anisotropic and space-time dependent. 

2.- The time dependence of e(T) follows from a particular application of
the equivalence principle for short intervals of time. It guarantees
that at this level of approximation, as far as electromagnetism is
concerned, the cosmological medium behaves as vacuum behaves in local
physics.

3.- We have assumed that $\alpha$ is a true constant. This is as much
conservative as an assumption as it is reckless to take for granted that
it is a function of time. It must be understood that what we claim
is that in our general framework, where a few constants $x$ have indeed a
time dependence of the form $x(T)=x_0F(T)^n$, in particular $n=0$ for
$\alpha$, but also for $r_0$ and $\lambda_c$\,\footnote{or the Bohr radius of an
hydrogen atom
$a_0=\epsilon(T)h(T)^2/(\pi m(T)e(T)^2)$}.
But this is, we
claim, the ground level behaviour and we do not exclude that $\alpha$ as
any other of the quantities that we have considered can have slightly
different behaviours if perturbations of the background cosmological
models are necessary to describe more realistic ones.

Notice also that we could have assumed that $r_0$ and $\lambda_c$
were true constants and derive from this the time dependence of $h(T)$
and the true constancy of $\alpha$. This would emphasize the primacy of
the dimension length and the dependence of the dimension velocity as
being derived from measures of length and time.

4.-The time dependence of $m(T)$ follows from more indirect
considerations, and is actually a substitute for a frequent claim
about a time dependence of $G$, which follows from other theories but would
be contradictory with our approach.  

The fact that the time dependence of $c(T)$ and the derived dependences
of $h(T)$ and $m(T)$ leads to a different relationship beween the
massless and massive modes of quantized fields in a cosmological context
provides an example where, without entering into deep metrology problems,
accepting time dependence of fundamental constants can lead to
interesting new scenarios.

\section*{Acknowledgments}

I gratefully acknowledge the position of visiting professor to the
UPV/EHU that I have been holding while this paper was being prepared.

\end{document}